# Exploring IoT in Smart Cities: Practices, Challenges and Way Forward

**Kashif Ishaq[1*], Syed Shah Farooq[1]**
[1]School of Systems and Technology, University of Management and Technology, 54000, Lahore, Pakistan

**Corresponding Author Email:**
kashif.ishaq@umt.edu.pk

## Abstract:

The rise of Internet of things (IoT) technology has revolutionized urban living, offering immense potential for smart cities in which smart home, smart infrastructure, and smart industry are essential aspects that contribute to the development of intelligent urban ecosystems. The integration of smart home technology raises concerns regarding data privacy and security, while smart infrastructure implementation demands robust networking and interoperability solutions. Simultaneously, deploying IoT in industrial settings faces challenges related to scalability, standardization, and data management. This research paper offers a systematic literature review of published research in the field of IoT in smart cities including 55 relevant primary studies that have been published in reputable journals and conferences. This extensive literature review explores and evaluates various aspects of smart home, smart infrastructure, and smart industry and the challenges like security and privacy, smart sensors, interoperability and standardization. We provide a unified perspective, as we seek to enhance the efficiency and effectiveness of smart cities while overcoming security concerns. It then explores their potential for collective integration and impact on the development of smart cities. Furthermore, this study addresses the challenges associated with each component individually and explores their combined impact on enhancing urban efficiency and sustainability. Through a comprehensive analysis of security concerns, this research successfully integrates these IoT components in a unified approach, presenting a holistic framework for building smart cities of the future. Integrating smart home, smart infrastructure, and smart industry, this research highlights the significance of an integrated approach in developing smart cities.

## Introduction:

The advent of the IoT has brought about a revolutionary shift in the idea of smart cities, where the integration of IoT technology assumes a crucial role in the metamorphosis of urban landscapes into intelligent, interconnected, and sustainable environments. The realm of IoT in smart cities encompasses a diverse range of variables that contribute to the overall functionality and progress of these urban ecosystems (Madakam et al., 2015). Smart cities rely on a robust smart infrastructure that serves as the foundation, incorporating sectors like transportation, energy, utilities, and public services. With the integration of the IoT, sensors embedded in infrastructure components such as roads, buildings, and utilities facilitate the monitoring of vital aspects including traffic flow, energy consumption, waste management, and other key factors (Muhammed et al., 2018). The idea behind smart homes centers on utilizing IoT technology to improve residents' quality of life. Within these homes, IoT-enabled devices and systems provide effortless connectivity, automation, and control over different elements like lighting, temperature, security, and energy consumption. Through embracing these advancements, residents can enjoy enhanced comfort, convenience, and energy efficiency within their living spaces (Woo & Lim, 2015). The advent of IoT has sparked remarkable changes in the industrial sector of smart cities. Smart industry, also known as Industry 4.0, leverages IoT devices, data analytics, and automation to elevate productivity, efficiency, and safety in manufacturing processes. This integration of technologies enables optimized production, reduced downtime, and improved worker safety, leading to significant advancements in the industrial landscape (Li et al., 2021).

IoT is a framework that incorporates various gadgets and innovations, eliminating the need of human mediation. This empowers the limit of having shrewd (or more brilliant) urban communities all over the planet. Through facilitating various advances and permitting associations between them, the web of things has initiated the improvement of brilliant city frameworks for practical living, expanded solace and efficiency for residents (Syed et al., 2021). The IoT empowers physical objects to perceive, comprehend, communicate, and perform tasks by facilitating communication and information sharing among them. Through the utilization of pervasive and ubiquitous computing, connected devices, communication technologies, sensor networks, internet protocols, and applications, the IoT transforms ordinary objects into smart ones. This transformation occurs by leveraging the underlying technologies of the IoT, enabling objects to seamlessly interact, exchange data, and make informed decisions. Through harnessing these capabilities, the IoT brings about a paradigm shift in the functionality and intelligence of everyday objects, unlocking new possibilities for efficiency, automation, and connectivity (Al-Fuqaha et al., 2015). Various innovative technologies and strategies that underpin these models provide intelligent services to enhance performance and operations in healthcare, transportation, energy, education, and many other domains. Via leveraging these advancements, smart cities can achieve greater efficiency, sustainability, and improved quality of life for their residents (Jawhar et al., 2018). The rise of another sort of systems administration worldview which empowers actual items to convey with the web, known as the IoT has grabbed the eye of the exploration networks and the Data and Correspondence Innovation (DCI) industry. The quantity of the gadgets and information created by the gadgets in IoT are on an ascent, as per a conjecture the IoT may have 50 billion units by 2020 (Lawal et al., 2020).

However, we will explore the fundamental variables that encompass smart infrastructure, smart homes, and smart industries, shedding light on their importance and influence within the realm of IoT in smart cities, its challenges and practices as shown in Figure 1. In this paper, we will explore into the essential factors comprising smart infrastructure, smart homes, and smart industry, as well as the obstacles that arise when implementing IoT within the framework of smart cities and combining these components in one research. Through examining these key variables and associated challenges, we aim to gain a comprehensive understanding of the impact and significance of IoT in shaping the future of intelligent urban environments.

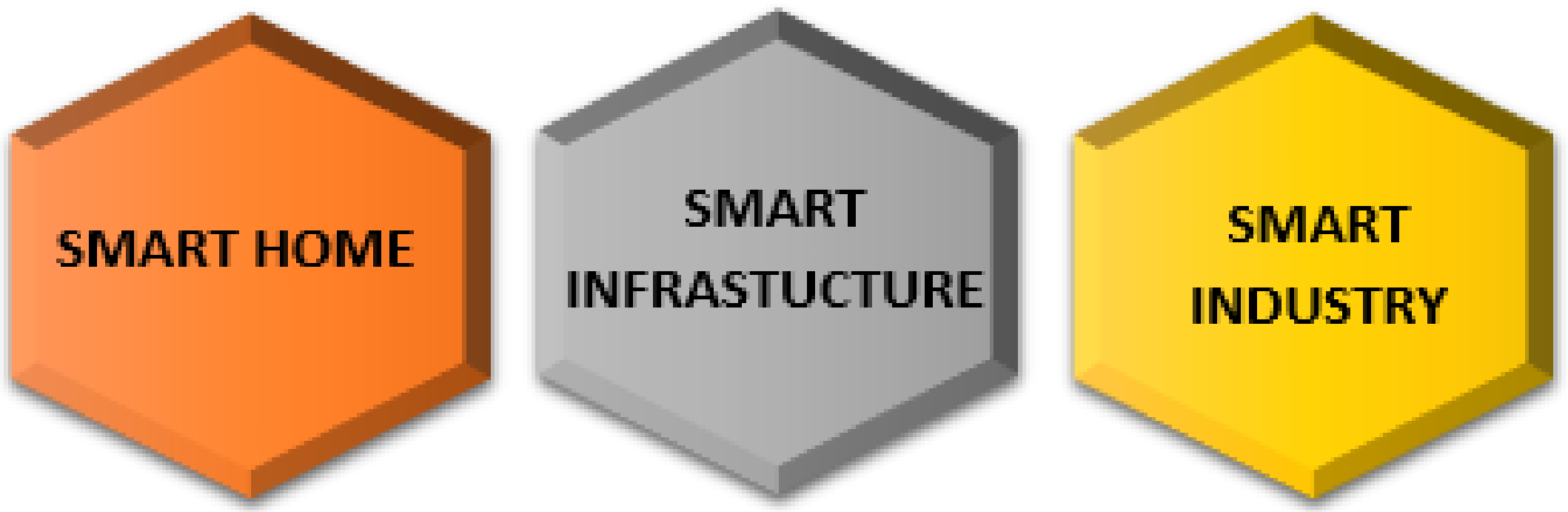

**Figure 1.** (Smart City Components)

**Background of the Study:**

Numerous studies have examined the function of IoT in various smart city domains. For instance, Smith (2020) states at how IoT devices and sensors are integrated into transportation systems, highlighting their potential to enhance mobility all around and enhance traffic management and public transportation route optimization. Li et al. (2021) demonstrated on the use of IoT-enabled solutions for energy management in smart cities, showing the efficiency of real-time monitoring, smart grid integration, and demand response mechanisms in achieving energy efficiency and sustainability goals. Researchers have looked into the opportunities and difficulties related to IoT implementation in smart cities in addition to specific domains. Ismagilova et al. (2020) states the major technological, security, privacy, and governance concerns that must be resolved for the successful implementation of IoT in smart cities. To overcome these obstacles and fully realize the potential benefits of IoT, they emphasized the significance of a solid infrastructure, data protection policies, and stakeholder collaboration. The study of Li et al. (2021) investigates into the IoT potential to enhance public security and safety in smart cities is another important study. For proactive threat detection and quick incident response, the study investigated the integration of IoT devices with surveillance systems, emergency response systems, and data analytics. It identified how IoT technology has the potential to improve situational awareness, lower crime rates, and guarantee citizen safety in environments like smart cities.

Furthermore, Sadhu et al., (2022) looked into the opportunities and problems related to the scalability and interoperability of IoT systems in smart cities, to ensure seamless communication and effective operation of various IoT devices and networks within urban environments, the study addressed issues relating to data integration, device connectivity, and standardization. In order to support the expansion and sustainability of smart cities, the findings highlighted the significance of scalable and interoperable IoT architectures.

These studies, along with others in the field, help us understand the many aspects of "IoT in Smart Cities." They also emphasize the issues that must be resolved for successful IoT implementation in smart city contexts. It also emphasizes how crucial it is to address technical, social, and regulatory issues in order to fully realize the advantages of IoT in building smarter, more sustainable urban environments. IoT technology serves as a foundation for additional study and the development of policy, allowing us to fully realize its potential for creating sustainable, interconnected urban environments.

**Table 1 Related Work Comparison**

| Reference | Title | Survey approach | Quality Assessment | Framework | Results | Data based on |
|---|---|---|---|---|---|---|
| (Al-Fuqaha et al., 2015) | Internet of Things: A Survey on Enabling Technologies, Protocols, and Applications | Informal | × | √ | √ | IEEE Xplore |
| (Lawal et al., 2020) | Security Analysis of Network Anomalies Mitigation Schemes in IoT Networks | Systematic Search | √ | √ | √ | IEEE Xplore |
| (Syed et al., 2021) | IoT in Smart Cities: A Survey of Technologies, Practices and Challenges | Informal | √ | √ | × | MDPI |
| (Jawhar et al., 2018) | Networking architectures and protocols for smart city systems | Systematic Search, Quality Assessment | × | × | √ | Springer Link |
| (Al-Badi et al., 2020) | Investigating Emerging Technologies Role in Smart Cities' Solutions | Systematic Search | √ | × | × | Springer Link |
| (Ferrag et al., 2020) | Security and Privacy for Green IoT-Based Agriculture: Review, Block chain Solutions, and Challenges | Informal | × | √ | √ | IEEE Xplore |
| This study | Exploring IoT in Smart Cities: Practices, Challenges and Way Forward | Systematic Search, Quality Assessment, Snowballing | √ | √ | √ | WoS Core Collection |

## Research Methodology:

This paper is about implementing IoT in smart cities, where it focused on integrating smart infrastructure, smart homes, and smart industry within a single smart city. It will also discuss the challenges associated with this integrated approach as shown in Figure 2.

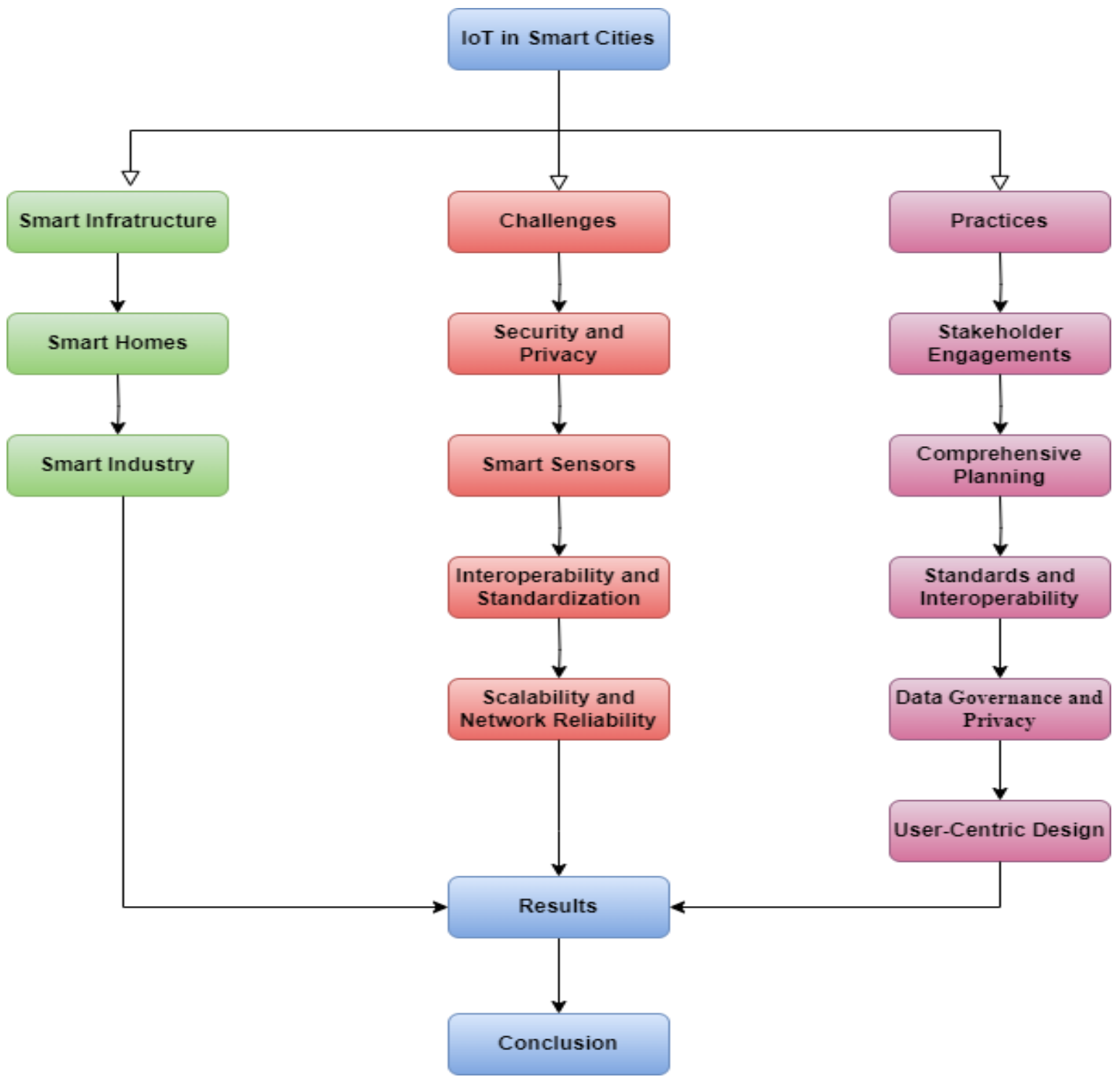

**Figure 2.** Flowchart of Methodology

## Research Questions:

| Table 2 Research Questions (RQs) | | |
|---|---|---|
| | **RQ Statement** | **Objectives and Motivation** |
| **RQ1:** | What are the high-quality publication channels for IoT in smart cities research, and which geographical areas have been targeting IoT in smart cities research over the years? | The objective of RQ1 is to search for high-quality research articles through major publications channels for IoT in smart cities including the geographical areas and publication, trended over the years. |
| **RQ2:** | What is the quality assessment of the relevant publications? | The quality assessment for the selected articles and the meta-information extracted useful statistics regarding framework/model, methodology, and result. |
| **RQ3:** | What are the key components and features of a smart infrastructure, and how does it contribute to creating sustainable and efficient cities? | IoT can lead to increased convenience and comfort for citizens. Through interconnected devices, residents can access various services remotely, such as smart home automation, intelligent transportation systems, and digital healthcare. This connectivity simplifies daily tasks, improves mobility, and promotes a more seamless and personalized living experience. |
| **RQ4:** | How can smart homes enhance the quality of life for residents by integrating various IoT devices and technologies and target IoT smart cities, its practices and challenges research?? | IoT brings numerous advantages to smart cities. By integrating various devices, sensors, and systems, IoT enables cities to gather and analyze vast amounts of data, leading to more efficient operations and enhanced services for residents. |
| **RQ5:** | How does the integration of smart industry solutions contribute to the overall development and efficiency of smart cities? | By integrating various IoT devices, smart homes offer numerous benefits that enhance residents' quality of life. E.g. Convenience and efficiency, energy management, safety and security, health and wellness, entertainment and connectivity. |
| **RQ6:** | What are the key principles and guiding practices for the successful implementation of IoT in smart cities? | These principles and practices can help ensure a smooth and effective deployment of IoT in smart cities. E.g. Collaboration and stakeholder engagement, comprehensive planning and strategy, standards and interoperability, data governance and privacy, user-centric design, scalability and flexibility, sustainability and resilience. |

## Inclusion and Exclusion Criteria:

### Inclusion Criteria:

The review paper centers around the topic of "IoT in Smart Cities," focusing on addressing specific research questions. The inclusion criteria for the reviewed papers encompass those published in journals or conferences from the period 2015 to 2022. Moreover, the review incorporates papers that discuss the IoT in Smart Cities at educational levels ranging from school, college, to university. The primary focus of these papers is to explore, develop, and discuss all possible ways in which IoT is applied in various domains, such as Smart Infrastructure, Smart Home, and Smart Industry.

### Exclusion Criteria:

Articles that were not written in an appropriate manner and did not discuss or focus on the IoT in Smart Cities, Smart Infrastructure, Smart Home, and Smart Industry within educational settings such as schools, colleges, and universities were excluded from the review. The aim was to concentrate on papers that specifically addressed and explored the applications of IoT in these domains within the context of educational institutions. The detailed selection process for including or excluding relevant articles as shown in Figure 3 and results shown in Table 3.

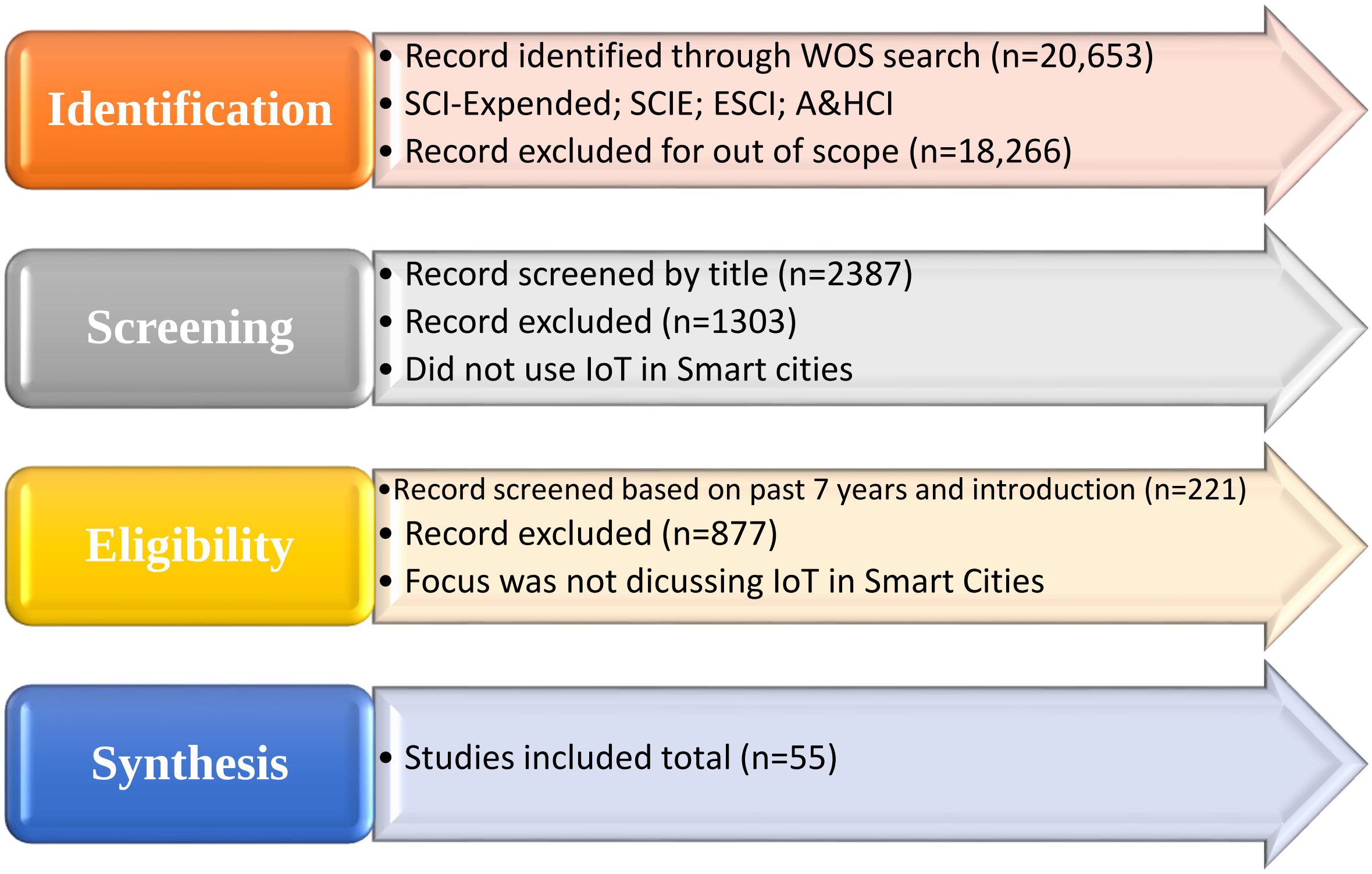

**Figure 3. Selection of Relevant Articles**

**Table 3. Results of Selection Phase**

| Phase # | Selection | Criteria | Indexes: SCI-EXPANDED, SSCI, A&HCI, ESCI |
|---|---|---|---|
| **1** | Search | Keywords (Figure) | 20,653 |
| **2** | Filtering | Title | 2,387 |
| **3** | Filtering | Abstract | 1084 |
| **4** | Filtering | Recent 7 Years | 656 |
| **5** | Filtering | Introduction | 221 |
| **6** | Inspection | Complete Article | 55 |

## Discussion of Research Questions:

In this section, we thoroughly examined and analyzed 55 primary research studies that were selected based on our research questions.

**RQ1: What are the high-quality publication channels for IoT in smart cities research, and which geographical areas have been targeting IoT in smart cities research over the years?**

The analysis of IoT in Smart Cities with the integration of all the elements in implementing, methods, content, and theoretical perspective was indeed challenging for researchers to implement in smart cities. Identifying suitable publication sites and conducting systematic research analysis based on meta-data in the IoT in Smart Cities field was necessary. This section involved gathering insightful information about research publication sources, types, years, grade level distribution, geographical dispersion, and distribution channel-wise distribution of selected studies to analyze IoT in Smart Cities research.

The examinations retrieved from the Web of Science (core collection) were presented annually, as shown in Table 4. **Detail of Publication by year**and Figure 4. The highest number of publications, twelve in total, were selected from the year 2020, indicating increased interest in developing IoT in Smart Cities. However, less interest in the IoT in Smart Cities, research was observed in the years 2016, 2018, 2021, and 2022.

**Table 4. Detail of Publication by year**

| Year | 2015 | 2016 | 2017 | 2018 | 2019 | 2020 | 2021 | 2022 |
|---|---|---|---|---|---|---|---|---|
| No. of publications | 7 | 5 | 10 | 5 | 6 | 12 | 4 | 6 |

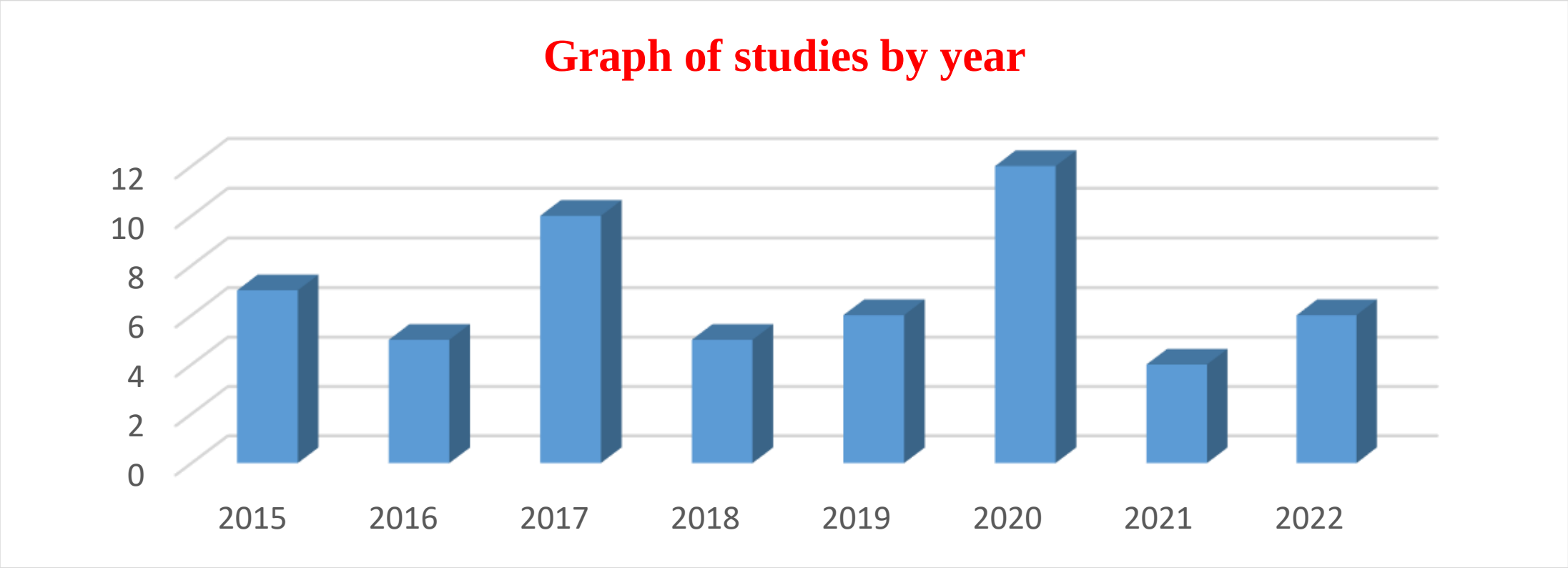

**Figure 4. Studies by year**

The information presented in Table 6 reveals that the majority of articles came from highly reputed journals listed in the Web of Science. Only one article was sourced from a well-regarded Conference. The "Computer and Education" journal ranked at the top, with seven papers selected, followed by the "Educational Technology and Society" journal with four articles.

| Table 5. Publications by geographic areas | | | |
|---|---|---|---|
| **Item no.** | **Sub-Content** | **Countries** | **Number of publications** |
| **1.** | Asia | China | 8 |
| | | Saudi Arabia | 3 |
| | | Iran | 2 |
| | | Malaysia | 1 |
| | | Sri Lanka | 4 |
| | | Bangladesh | 5 |
| | | Pakistan | 3 |
| | | Turkey | 4 |
| | | Sultanate of Oman | 3 |
| **2.** | Europe | UK | 6 |
| | | Norway | 2 |
| | | Italy | 1 |
| | | Finland | 1 |
| | | Netherlands | 2 |
| **3.** | North America | USA | 5 |
| | | Canada | 1 |
| | | Indiana | 1 |
| **4.** | Africa | South Africa | 2 |
| | | Morocco | 1 |
| **Total** | | | **55** |

Table 5 and **Error! Reference source not found.** display the geographically distributed investigations. The majority of the publications, 21 out of 55, originated from various Asian countries, whereas European countries contributed 12 studies. North American countries published seven investigations, while Africa and the Middle East had three and four studies published, respectively. Finally, the Indian Subcontinent and Muscat each had four and three studies published, respectively.

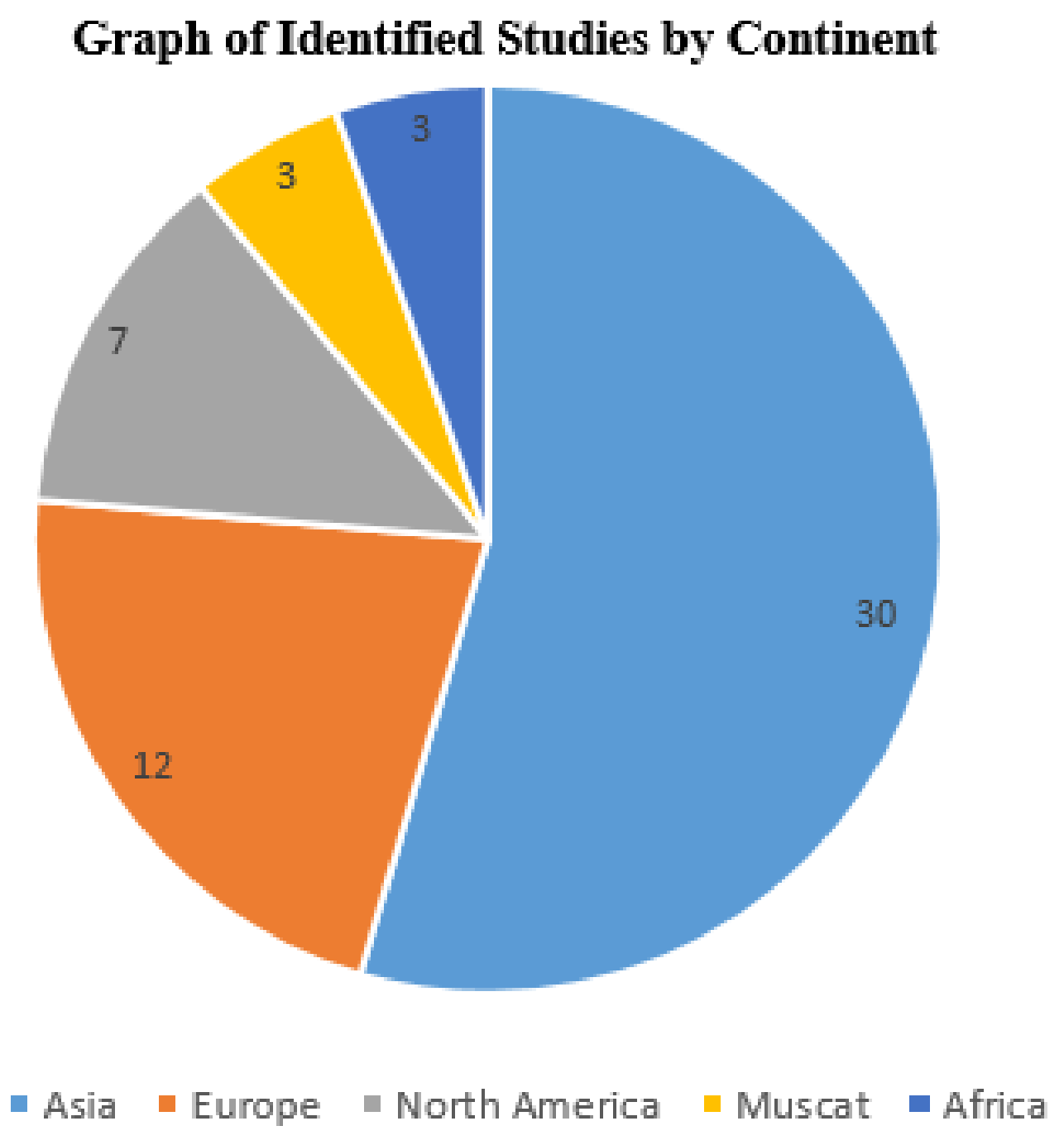

**Figure 5 (Graph of Identified Studies by Continent)**

| Item no. | Publication Source | Channel | No. of Articles |
|---|---|---|---|
| **Table 6. Publications sources** | | | |
| **1.** | Education and Information Technologies | Journal | 15 |
| **2.** | Computers & Education | Journal | 11 |
| **3.** | Computer-Assisted Language Learning | Journal | 7 |
| **4.** | Journal of Asia TEFL | Journal | 2 |
| **5.** | International Journal of Advanced Computer Science and Applications | Journal | 2 |
| **6.** | Interactive Learning Environments | Journal | 1 |
| **7.** | Journal of Educational Computing Research | Journal | 1 |
| **8.** | International Journal of Mobile and Blended Learning | Journal | 1 |
| **9.** | Education Sciences | Journal | 1 |
| **10.** | British Journal of Educational Technology | Journal | 1 |
| **11.** | International Review of Research in Open and Distributed Learning | Journal | 1 |
| **12.** | Interactive Technology and Smart Education | Journal | 1 |
| **13.** | Technology Pedagogy and Education | Journal | 1 |
| **14.** | Journal of Educational Computing Research | Journal | 1 |
| **15.** | Journal of Information Technology Education-Research | Journal | 1 |
| **16.** | Journal of Computers in Education | Journal | 1 |
| **17.** | International Journal of Instruction | Journal | 1 |
| **18.** | 2019 International Conference on Computer and Information Sciences (ICCIS) | Conference | 1 |
| **19.** | 2018 International Conference on Current Trends towards Converging Technologies (ICCTCT) | Conference | 1 |
| **20.** | 2019 International Conference on Innovative Computing (ICIC) | Conference | 1 |
| **21.** | 2020 IEEE 20th International Conference on Advanced Learning Technologies (ICALT) | Conference | 1 |
| **22.** | 2022 IEEE 21th International Conference on Advanced Learning Technologies (ICALT) | Conference | 1 |

## RQ2: What is the quality assessment of the relevant publications?

As per the criteria defined in "Research Methodology," each completed review's QA score was assigned with QA scores ranging from 4 to 8, and scores below 4 being excluded. The QA score proved helpful for analysts studying the IoT in smart cities, as it facilitated the selection of relevant studies while addressing its applications and challenges. Articles published in Q1 journals generally scored the highest, with four scores from less recognized journals that were still relevant to the topic. Out of 55 reviews, a total of 24 scored the maximum (8), indicating fulfillment of all QA standards, while (9) reviews scored four out of 55, which is the lowest in QA. **Error! Reference source not found.** provides an overview of the overall classification output and quality assessment (QA) of the finalized studies and **Error! Reference source not found.** showed the overall quality assessment score. The studies were categorized based on five factors: empirical type/method, research type, and method. The research types fell into four categories: Evaluation framework, Evaluation research, Solution proposal, and Review. To enhance their quality standards, studies that analytically validated their results through statistical analysis, experiments, surveys, or case studies were awarded scores. In category (c) of the quality assessment criteria, only 6 out of 26 reviews did not present empirical results and thus received zero scores. For category (d) of the quality assessment criteria, only two studies scored zero, while the remaining studies received higher scores, indicating their credibility as reliable sources. Six studies received the lowest score among all the assessed studies.

**Table 7 Quality Assessment**

| Ref. | Classification | | | | | Quality assessment | | | | |
|---|---|---|---|---|---|---|---|---|---|---|
| | P. channel | Publication year | Research type | Empirical type/method | Methodology | (a) | (b) | (c) | (d) | Score |
| (Al-Badi et al., 2020) | Research Journal | 2020 | Evaluation Framework | Survey | Formulation of conceptual model | 1 | 1 | 2 | 4 | 8 |
| (Bellini et al. (2022) | Research Journal | 2022 | Evaluation Framework | Experiment | Statistical Analysis | 1 | 0 | 1 | 3 | 5 |
| (Misbahuddin et al., 2015) | Research Journal | 2015 | Evaluation Research | Survey | Statistical Analysis | 1 | 0 | 1 | 4 | 6 |
| (Papadakis et al., 2021) | Research Journal | 2021 | Evaluation Research | Survey | Personalized recommendation-based approach | 1 | 2 | 1 | 4 | 8 |
| (Esmaeilian et al., 2018) | Research Journal | 2018 | Evaluation Research | Survey | Statistical Analysis | 1 | 2 | 1 | 2 | 6 |
| (Cui et al., 2018) | Research Journal | 2018 | Solution Proposal | Experiment | Personalized recommendation-based approach | 1 | 2 | 1 | 4 | 8 |
| (Zhang et al., 2021) | Research Journal | 2021 | Evaluation Research | Survey | Questionnaire | 1 | 1 | 2 | 4 | 8 |
| (Woo & Lim, 2015) | Research Journal | 2015 | Evaluation Research | Survey | Interview and Observation | 1 | 1 | 1 | 2 | 5 |
| (Madakam et al., 2015) | Research Journal | 2015 | Evaluation Research | Experiment | Statistical Analysis | 1 | 1 | 1 | 3 | 6 |
| (Botta et al., 2016) | Research Journal | 2016 | Solution Proposal | Survey | Formulation of conceptual model | 2 | 1 | 2 | 3 | 8 |
| (Herath & Mittal, 2022) | Research Journal | 2022 | Evaluation Research | Survey | Statistical Analysis | 1 | 1 | 3 | 2 | 7 |
| (Jawhar et al., 2018) | Research Journal | 2018 | Evaluation Research | Experiment + Survey | Questionnaire, Semi-Structured Interview and Observation | 1 | 2 | 1 | 4 | 8 |
| (Sosunova & Porras, 2022) | Research Journal | 2022 | Solution Proposal | Survey | Questionnaire | 1 | 3 | 0 | 3 | 7 |
| (Ystgaard et al., 2023) | Research Journal | 2023 | Evaluation Research | Survey | Statistical Analysis | 1 | 2 | 1 | 4 | 8 |
| (Al-Turjman & Malekloo, 2019) | Research Journal | 2019 | Evaluation Research | Survey | Statistical Analysis | 1 | 0 | 2 | 3 | 6 |
| (Lau et al., 2015) | Research Journal | 2015 | Solution Proposal | Experiment | Formulation of conceptual model | 1 | 0 | 3 | 0 | 4 |
| (Damadam et al., 2022) | Research Journal | 2022 | Solution Proposal | Survey | Statistical Analysis | 1 | 0 | 0 | 4 | 5 |
| (Hashem et al., 2016) | Research Journal | 2016 | Evaluation Research | Survey | Statistical Analysis | 1 | 2 | 1 | 4 | 8 |
| (Islam et al., 2020) | Research Journal | 2020 | Evaluation Research | Survey | Statistical Analysis | 1 | 0 | 2 | 3 | 6 |
| (Ejaz et al., 2017) | Research Journal | 2017 | Evaluation Research | Survey | Statistical Analysis | 1 | 2 | 0 | 4 | 7 |
| (Donbosco & Chakraborty, 2021) | Research Journal | 2021 | Evaluation Research | Experiment + Survey | Questionnaire and Observation | 1 | 0 | 1 | 3 | 5 |
| (Costa et al., 2022) | Research Journal | 2022 | Evaluation Research | Survey | Survey and Interview | 1 | 2 | 0 | 2 | 5 |
| (Ferrag et al., 2020) | Research Journal | 2020 | Evaluation Research | Survey | Statistical Analysis | 1 | 2 | 1 | 0 | 4 |
| (Khan et al., 2019) | Research Journal | 2019 | Evaluation Research | Survey | Interview and Observation | 1 | 2 | 1 | 2 | 6 |
| (Bhutta & Ahmad, 2021) | Research Journal | 2021 | Evaluation Research | Survey | Statistical Analysis | 1 | 0 | 1 | 2 | 4 |
| (Herath & Mittal, 2022) | Research Journal | 2022 | Evaluation Research | Survey | Statistical Analysis | 1 | 1 | 0 | 2 | 4 |
| (Sehrawat & Gill, 2019) | Research Journal | 2019 | Evaluation Research | Survey | Statistical Analysis | 1 | 1 | 3 | 2 | 7 |
| (Samuel, 2016) | Research Journal | 2016 | Evaluation Research | Survey | Statistical Analysis | 1 | 2 | 2 | 2 | 7 |
| (Sharma et al., 2021) | Research Journal | 2021 | Evaluation Research | Survey | Statistical Analysis | 1 | 1 | 0 | 1 | 4 |
| (Pieroni et al., 2018) | Research Journal | 2018 | Evaluation Research | Survey | Statistical Analysis | 1 | 2 | 3 | 2 | 8 |
| (Muthuramalingam et al., 2019) | Research Journal | 2019 | Evaluation Research | Survey | Statistical Analysis | 1 | 0 | 1 | 2 | 4 |
| (Donbosco & Chakraborty, 2021) | Research Journal | 2021 | Evaluation Research | Experiment + Survey | Statistical Analysis | 1 | 2 | 3 | 2 | 8 |
| (Simonofski et al., 2017) | Research Journal | 2017 | Evaluation Research | Survey | Statistical Analysis | 1 | 1 | 3 | 2 | 7 |
| (Mahmood et al., 2018) | Research Journal | 2018 | Evaluation Research | Survey | Statistical Analysis | 1 | 2 | 2 | 2 | 7 |

## Table 8. Quality Assessment Score

| References | Score | Total |
|---|---|---|
| (Al-Badi et al., 2020) (Papadakis et al., 2021) (Zhang et al., 2021) (Cui et al., 2018) (Botta et al., 2016) (Jawhar et al., 2018) (Ystgaard et al., 2023) (Hashem et al., 2016) (Pieroni et al., 2018) (Donbosco & Chakraborty, 2021) | 8 | 10 |
| (Herath & Mittal, 2022) (Ejaz et al., 2017) (Sosunova & Porras, 2022) (Sehrawat & Gill, 2019) (Samuel, 2016) (Simonofski et al., 2017) (Mahmood et al., 2018) | 7 | 7 |
| (Misbahuddin et al., 2015) (Esmaeilian et al., 2018) (Madakam et al., 2015) (Al-Turjman & Malekloo, 2019) (Islam et al., 2020) (Khan et al., 2019) | 6 | 6 |
| (Bellini et al., 2022) (Woo & Lim, 2015) (Damadam et al., 2022) (Donbosco & Chakraborty, 2021) (Costa et al., 2022) | 5 | 5 |
| (Lau et al., 2015) (Ferrag et al., 2020) (Bhutta & Ahmad, 2021) (Herath & Mittal, 2022) (Sharma et al., 2021) (Muthuramalingam et al., 2019) | 4 | 6 |

**RQ3: What are the key components and features of a smart infrastructure, and how does it contribute to creating sustainable and efficient cities?**

Smart infrastructure is a connected system of diverse components and technologies that leverage data and advanced communication to enhance city functioning as shown in Figure 6. It comprises several key features:

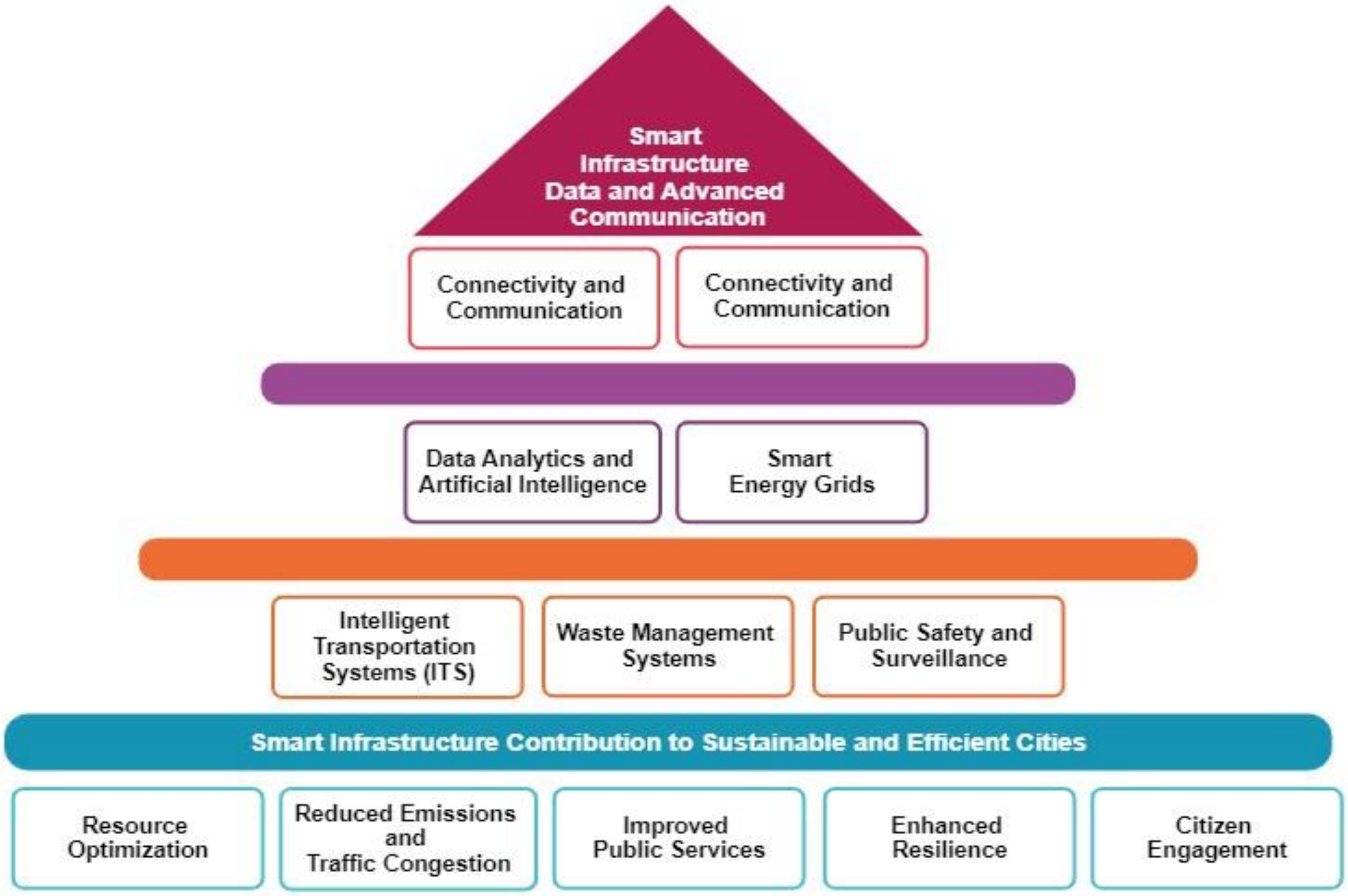

**Figure 6. Key Components and Features of Smart Infrastructure**

**Sensors and IoT Devices:** Smart infrastructure deploys a network of sensors and IoT devices across the city. These sensors collect real-time data on various aspects like traffic flow, air quality, energy consumption, waste management, and water usage (Sehrawat & Gill, 2019).

**Connectivity and Communication:** A reliable communication network is crucial for smart infrastructure. This network enables seamless connectivity and data exchange between components, allowing real-time monitoring and analysis (Samuel, 2016).

**Data Analytics and Artificial Intelligence:** Data from sensors and devices undergoes processing and analysis using data analytics and artificial intelligence (AI). AI algorithms derive insights, patterns, and trends from the data, aiding better decision-making and city service optimization (Sharma et al., 2021).

**Smart Energy Grids:** Smart infrastructure integrates intelligent energy grids that efficiently monitor and manage energy distribution. These grids can incorporate renewable energy sources, manage demand, and reduce energy wastage, promoting sustainability (Pieroni et al., 2018).

**Intelligent Transportation Systems (ITS):** Smart cities incorporate ITS to enhance mobility and reduce congestion. ITS includes technologies like real-time traffic monitoring, smart traffic lights, and advanced public transportation systems, boosting transportation efficiency and reducing carbon emissions (Muthuramalingam et al., 2019).

**Waste Management Systems:** Smart infrastructure adopts smart waste management systems that optimize collection schedules based on real-time data. This reduces operational costs and fosters a cleaner environment (Donbosco & Chakraborty, 2021).
**Public Safety and Surveillance:** Smart infrastructure includes advanced surveillance and public safety systems like smart CCTV cameras, emergency response systems, and early warning systems, enhancing city safety and resilience (Papadakis et al., 2021).

**Smart infrastructure contribution:**

Smart infrastructure is crucial to build sustainable and efficient cities. It enhances resource utilization, service delivery, and residents' quality of life by harnessing cutting-edge technologies and the Internet of Things (IoT). Smart infrastructure empowers city officials with real-time data insights through interconnected sensors and devices, enabling data-driven decision-making and proactive management of urban systems. The integration of intelligent transportation, efficient energy distribution, and optimized waste management is facilitated by the seamless connectivity and automation offered by smart infrastructure, leading to reduced carbon emissions and resource waste.
**Resource Optimization:** Data-driven insights enable cities to optimize resource usage, from energy consumption to waste management, leading to greater sustainability and cost savings (Esmaeilian et al., 2018).
**Reduced Emissions and Traffic Congestion:** Advanced transportation technologies reduce congestion and emissions, promoting environmental sustainability (Muthuramalingam et al., 2019).
**Maintenance and Asset Management:** IoT sensors enable predictive maintenance for city assets, such as bridges, roads, and public facilities. Smart infrastructure allows proactive identification of potential issues, leading to cost-effective and timely maintenance, prolonging the lifespan of critical assets (Heaton & Parlikad, 2019).
**Improved Public Services:** Predictive maintenance and real-time monitoring of critical systems enhance public services' reliability, improving residents' quality of life (Papadakis et al., 2021).
**Enhanced Resilience:** Real-time monitoring and data analysis improve cities' resilience during emergencies and natural disasters (Costa et al., 2022).
**Environmental Monitoring and Sustainability Initiatives:** Smart infrastructure includes environmental monitoring solutions that track air and water quality, noise levels, and other ecological parameters. These initiatives help identify environmental issues, allowing cities to implement targeted sustainability measures and improve overall environmental health (Perera et al., 2014).
**Citizen Engagement:** Smart infrastructure fosters transparency and citizen participation, allowing residents to actively shape their urban environment (Simonofski et al., 2017).
**Smart Street Lighting:** Smart infrastructure extends to street lighting systems that are equipped with IoT sensors and adaptive controls. These smart streetlights can adjust their brightness based on real-time conditions, such as ambient light levels and pedestrian activity, reducing energy consumption and light pollution (Parkash & Rajendra, 2016).
In conclusion, smart infrastructure, with its key components and data-driven approach, creates sustainable, efficient, and livable cities by optimizing resources, enhancing mobility, improving public services, and fostering citizen engagement.

**RQ4: How can smart homes enhance the quality of life for residents by integrating various IoT devices and technologies and target IoT smart cities, its practices and challenges research?**

Smart homes leverage the power of IoT to create an interconnected ecosystem of devices and technologies within a residential setting. Integrating various IoT devices, smart homes offer numerous benefits that enhance residents' quality of life as shown in Figure **8**7. Here are some ways in which smart homes accomplish this:

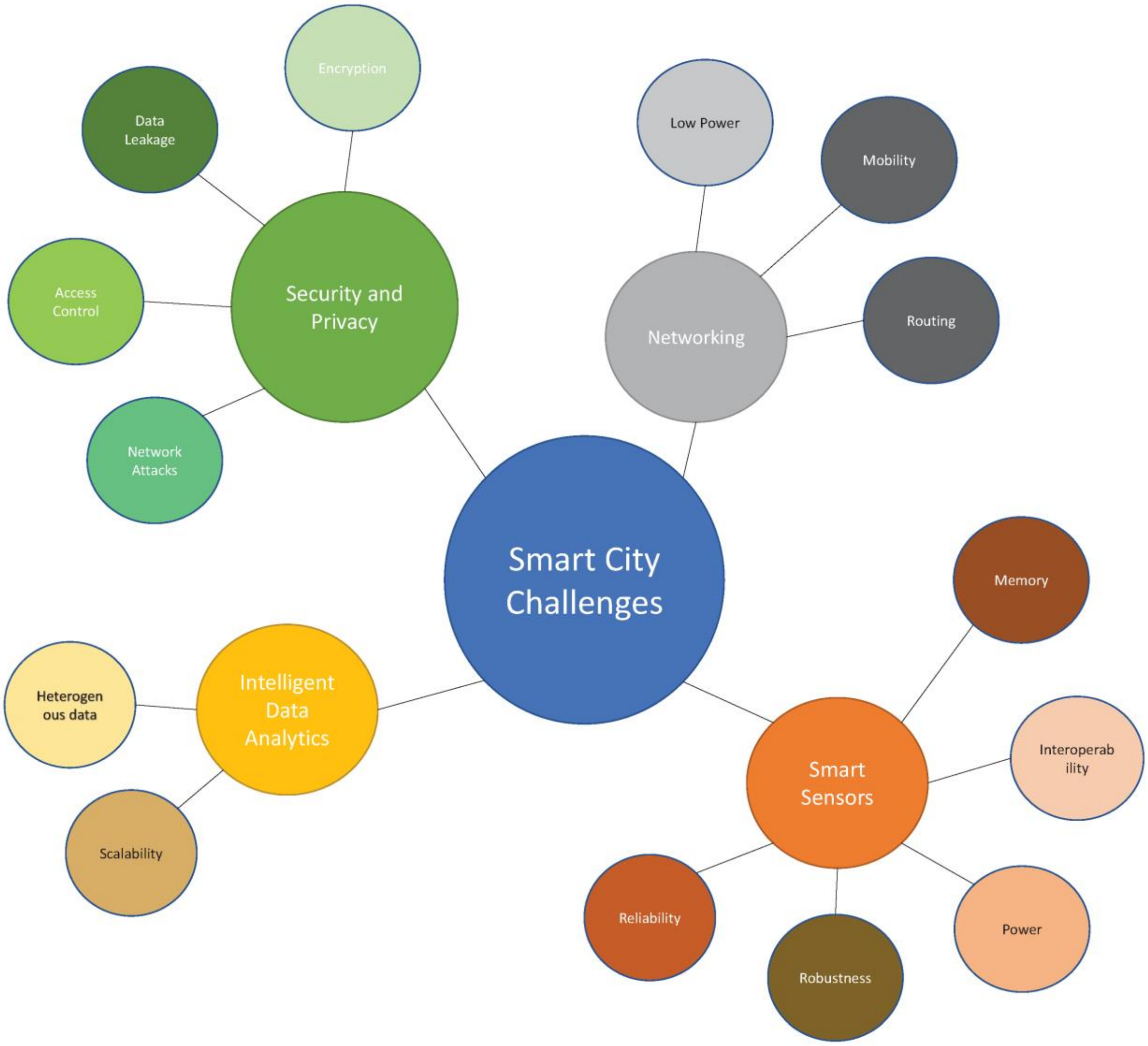

**Figure 7. Challenges for IoT in Smart Cities**

**Convenience and Efficiency:** Smart homes provide residents with greater convenience and efficiency in managing their daily tasks. Through IoT integration, residents can automate and control various devices such as lighting, thermostats, appliances, and security systems using voice commands or mobile applications. This level of automation simplifies routine activities, saves time, and improves overall efficiency (Hui et al., 2017).

**Energy Management:** IoT-enabled devices in smart homes allow for intelligent energy management. Residents can monitor and control energy consumption, optimize heating and cooling systems based on occupancy, and receive real-time energy usage data. These features promote energy efficiency, cost savings, and environmental sustainability (Ejaz et al., 2017).

**Safety and Security:** Smart homes offer advanced security features through IoT integration. Residents can monitor their homes remotely using security cameras, receive notifications for suspicious activities, and control access through smart locks. Integration with smoke detectors, carbon monoxide sensors, and water leak detectors enhances safety and provides early warnings in case of emergencies (Cui et al., 2018).

**Health and Wellness:** IoT devices can contribute to residents' health and wellness by monitoring and managing various aspects. Smart wearables, connected medical devices, and health monitoring systems can track vital signs, sleep patterns, and physical activity. Residents can receive personalized insights, reminders, and alerts to support their well-being (Islam et al., 2020).

**Entertainment and Connectivity:** Smart homes integrate entertainment systems, such as smart TVs, audio systems, and streaming devices, providing seamless connectivity and immersive experiences. Residents can access their favorite content, playlists, and online services effortlessly, enhancing their entertainment options and connectivity within the home (Mahmood et al., 2018).

**Accessibility and Aging in Place:** Smart homes can greatly benefit individuals with disabilities or older adults by providing assistive technologies and improving accessibility. IoT devices, such as voice-controlled assistants, smart lighting, and automated systems, enable easier interaction with the environment, allowing residents to age in place comfortably and independently (Fattah et al., 2017).

## IoT smart cities key challenges in implementation:

**Interoperability:** IoT involves integrating diverse devices, systems, and platforms from different vendors and manufacturers. Ensuring interoperability and seamless communication between these components can be a challenge. Standardization efforts are necessary to establish common protocols and interfaces that enable devices to work together effectively (Al-Badi et al., 2020).

**Scalability:** Smart cities encompass a large-scale deployment of IoT devices and systems. Scaling up the infrastructure and managing a vast number of devices pose challenges in terms of network capacity, data management, and system integration. Planning for scalability from the beginning is crucial to avoid bottlenecks and ensure smooth operations (Qian et al., 2019).

**Data Security and Privacy:** The vast amount of data generated by IoT devices raises concerns about data security and privacy. Protecting sensitive information from unauthorized access, ensuring secure communication, and implementing robust data protection measures are essential. Additionally, clear policies and regulations must address privacy concerns to build trust among residents and stakeholders (Ferrag et al., 2020).

**Infrastructure and Connectivity:** Implementing IoT in smart cities requires robust infrastructure and reliable connectivity. Ensuring widespread coverage of communication networks, including high-speed internet, cellular networks, and low-power wide-area networks, is crucial. Adequate infrastructure investment and network expansion are necessary to support the connectivity demands of IoT devices (Lawal et al., 2020).

**Cost and Funding:** Deploying IoT infrastructure and maintaining smart city initiatives can be expensive. The costs include device procurement, network deployment, data management systems, and ongoing maintenance. Securing funding and developing sustainable business models are crucial challenges in implementing IoT in smart cities (Perera et al, 2014).

**Regulatory and Legal Frameworks:** The rapid development of IoT technologies often outpaces the regulatory and legal frameworks governing their use. Smart city initiatives need clear regulations addressing data ownership, privacy, liability, and cybersecurity (Ismagilova et al., 2020). Establishing appropriate governance models and collaboration between public and private sectors is essential to navigate legal challenges.

## IoT Based Devices and Technologies in Smart Home:

| Table 7. Smart Home Technologies/Devices | | | | |
|---|---|---|---|---|
| **Reference** | **Description** | **Technology** | **IoT Device/Sensors** | **Limitation** |
| (Poongodi et al., 2021) | Patients' health can be checked remotely by Doctors, and by Patients for remedies. | Smart Health Monitoring System | SPO2, Temperature body position GSR<br>(LM35) & Heart Beat Rate Sensor<br>ECG Sensor<br>Glucometer EMG | Detect patients abnormal conditions, alerting doctors, and provide health services but the connection is interrupted. |
| (Aldegheishem et al., 2022) | Reduces the consumption of water resources as well as improves crop quality. | Monitoring and Irrigation of an Urban Garden | DHT22<br>MQ135<br>SHT10<br>BH1750<br>GP2YA21 | Need proper monitoring system in which the data and controlling of the devices should be controlled remotely. |
| (Shah & Mishra, 2016) | Monitoring pollution levels and warning people. | Smart Urban Climate Monitoring System | (DHT22)<br>(BMP180)<br>(BH1750)<br>(MQ7)<br>(MQ135) | The information which is delivered to the user is not accurate sometimes due to internet connection issues. |
| (Stolojescu-Crisan et al., 2021) | Smart Home appliances, lighting, heating, air conditioning, TVs, computers, entertainment audio & video systems, security, and camera systems. | Smart Home Automation System Appliances using Multiple Sensors | Raspberry PI<br>Node MCU<br>Arduino MEGA<br>ESP8266<br>Raspberry PI | The capability of communicating with one another and can be controlled remotely by a time schedule was interrupted and many bugs and errors occur. |
| (Kumar et al., 2018) | Detect patient's abnormal conditions, alerting doctors, and provide health services. | Intelligent Health Care Monitoring System | SPO AND ECG<br>Airflow<br>Temperature<br>Body Position GSR | Limitations of the Smart Healthcare Monitoring System include potential challenges in ensuring the compatibility of diverse medical devices and data sources, as well as addressing privacy concerns related to the collection and storage of sensitive patient information. |
| (Khandelwal et al., 2018) | Personal security analyzes the severity of crimes against women, reducing the rate of harassment. | Smart People Safety System | Heart Beat Sensor<br>(LM35)<br>GPS (EM-506) | Limitations of the Smart People Safety System encompass potential false alarms due to the complexity of recognizing genuine threats from various environmental factors. The system's accuracy heavily relies on sensor technology, which might be affected by external conditions. |
| (Khanna & Anand, 2016) | Online booking and improve parking facilities. | Smart Parking System | Raspberry Pi<br>(HC-SRO4)<br>Wi-Fi-ESP8266 | Limitations of the Smart Parking System include potential inaccuracies in detecting small vehicles or motorcycles, leading to misrepresented occupancy data. Network connectivity issues might result in |
| | | | | delays in real-time data updates, impacting the system's effectiveness in guiding drivers. |
| (Attri et al., 2016) | Delay reducing and enhance the response time. | Traffic Signal Preemption (TSP) System | Arduino (ATmega2560)<br>Inbuilt GPS of Smartphone<br>Wi-Fi module IEEE 802.11 | Limitations of the Traffic Signal Preemption (TSP) System include potential conflicts with emergency vehicles from different jurisdictions, causing coordination challenges. The system's effectiveness can be compromised by heavy traffic scenarios, as TSP might disrupt traffic flow and lead to congestion. |

In our interconnected world, smart technologies have revolutionized various aspects of urban living. Smart Health facilitates advanced healthcare delivery through wearable devices and remote monitoring, enhancing individual well-being. Monitoring systems for urban gardens employ IoT sensors to optimize irrigation and nurture plant growth, ensuring a greener and sustainable environment. The Smart Urban Climate Monitoring System leverages technology to keep track of weather patterns, air quality, and temperature in cities, aiding in climate-conscious urban planning. Meanwhile, Smart Home Automation Systems bring convenience and energy efficiency to households by using multiple sensors to control appliances and devices. These innovations collectively contribute to a smarter, healthier, and more sustainable urban lifestyle.

The Intelligent Health Care Monitoring System employs cutting-edge wearable devices and data analytics to revolutionize patient care by enabling remote health monitoring and enhancing medical outcomes. The Smart People Safety System harnesses technology to ensure citizen safety through real-time monitoring, yet faces challenges regarding sensor accuracy and user adoption. Meanwhile, the Smart Parking System optimizes urban mobility by providing real-time parking information, alleviating congestion and enhancing user convenience. The Traffic Signal Preemption (TSP) System strives to streamline emergency vehicle navigation, though its efficiency can be hindered by coordination issues and traffic dynamics. Collectively, these systems underscore the potential of technology to address urban complexities while underscoring the significance of seamless integration and addressing inherent limitations as shown in Table 7.

**RQ5: How does the integration of smart industry solutions contribute to the overall development and efficiency of smart cities?**

Integrating smart industry solutions into smart cities plays a significant role in their overall development and efficiency. Smart industry solutions leverage advanced technologies and data-driven approaches to enhance industrial processes, supply chains, and manufacturing operations. When integrated into smart cities, these solutions offer several key benefits shown in Figure **8**8:

**Key components for best solutions:**

**Improved Industrial Efficiency:** Smart industry solutions optimize industrial processes through automation, real-time monitoring, and predictive analytics. Through optimizing resource usage, reducing downtime, and streamlining operations, industries become more efficient and productive, leading to economic growth and increased competitiveness (Lom et al., 2016).

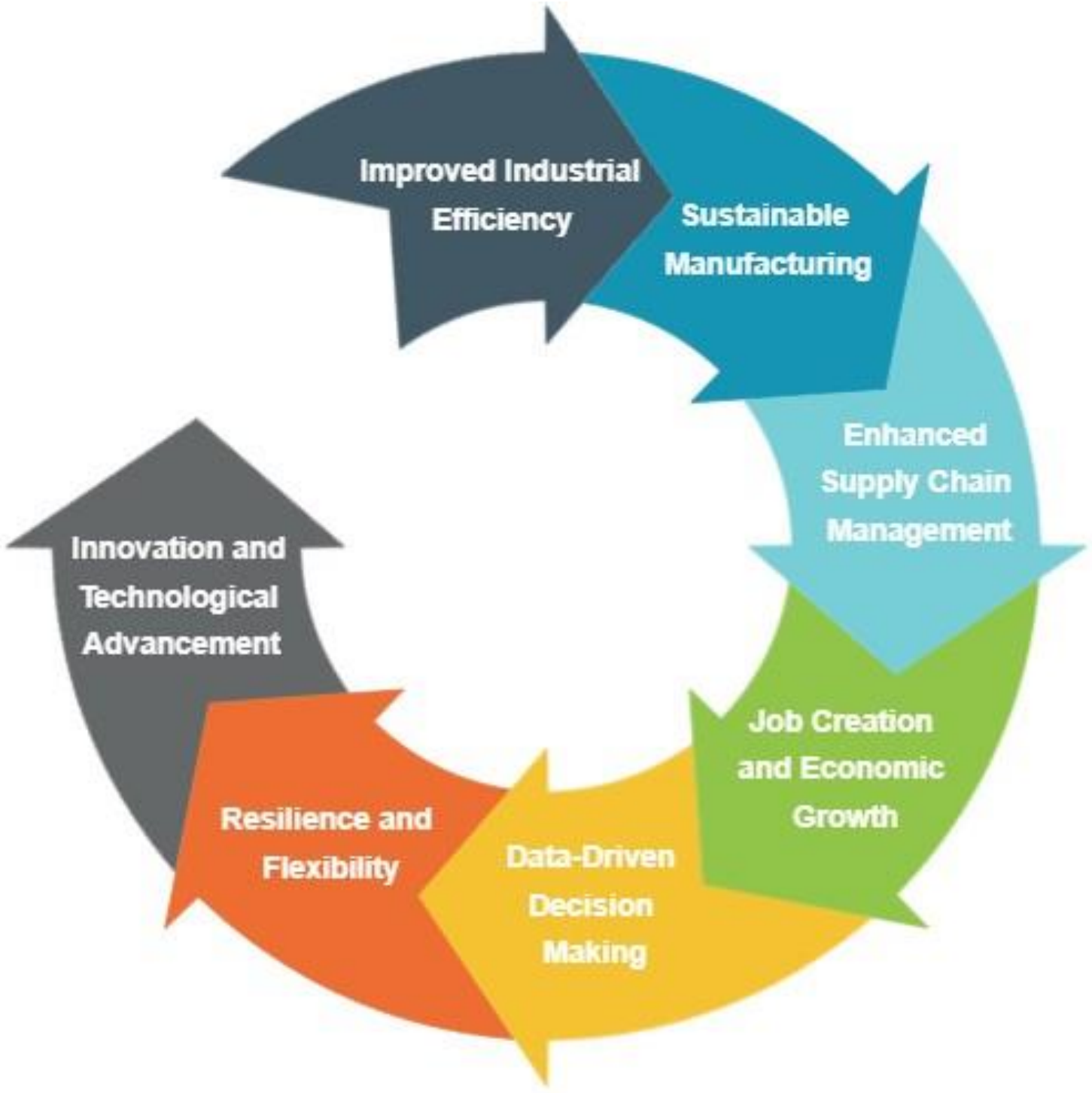

**Figure 8. Integration of Smart Industry Solutions**

**Sustainable Manufacturing:** Integrating smart industry solutions promotes sustainable manufacturing practices. Energy-efficient production, waste reduction, and resource optimization contribute to the city's overall sustainability goals, reducing the environmental impact of industrial activities (Syed et al., 2021).

**Enhanced Supply Chain Management:** Smart industry solutions provide better visibility and control over supply chains. Real-time tracking, data analytics, and automation streamline logistics, reducing delays and improving the efficiency of goods transportation and distribution within the city (Nagarajan et al., 2022).

**Job Creation and Economic Growth:** As industries become more efficient and competitive, they create new job opportunities and foster economic growth. Smart industry initiatives attract investments and foster innovation, leading to a vibrant business ecosystem within the city.

Seamless Integration with Smart Infrastructure: Integrating smart industry solutions with the city's existing smart infrastructure creates a synergistic effect. Data sharing and collaboration between industries and city services enable better resource management, supply chain optimization, and improved delivery of goods and services (Bonte, 2018).

**Data-Driven Decision Making:** Smart industry solutions generate vast amounts of data that can be harnessed to make informed decisions. The integration of this data with the broader city data ecosystem empowers city officials and industrial stakeholders with valuable insights, enabling data-driven planning, policy-making, and resource allocation (Andronie et al., 2021).

**Resilience and Flexibility:** Smart industry solutions enhance industrial resilience by enabling swift responses to disruptions and crises. Real-time data and predictive analytics help identify potential issues before they escalate, allowing proactive measures to be taken. This adaptability and flexibility contribute to the overall resilience of the smart city (Singh et al., 2020).

**Innovation and Technological Advancement:** The integration of smart industry solutions encourages a culture of innovation within the city. Industries adopting cutting-edge technologies and practices drive technological advancement, which spills over into other sectors and contributes to the overall development of the smart city (Lom et al., 2016).

## IoT Based Devices and Technologies in Smart Industries:

**Table 8 Smart Industries Solutions/Devices**

| Reference | Description | Industry | Sensors | Limitation | Solutions |
|---|---|---|---|---|---|
| (Nirde et al., 2017) | Maintain the city clean and keep the people away from diseases. | Smart Solid Waste Monitoring and Collection | (FSR 402)<br>(HC-SRO4)<br>PIR<br>GPS (EM-506) | The communication method is GSM or GPRS which is not reliable and accurate on these devices. | Solutions for the Smart Solid Waste Monitoring and Collection system involve using predictive analytics to optimize waste collection routes, reducing fuel consumption and emissions. Integrating smart bins with sensors can provide real-time data on fill levels. |
| (Dheena et al., 2017) | Energy saving and maintenances costs. | Smart Street Light Management System | LDR,<br>DHT11<br>Wi-Fi-ESP8266 | It had better be Web-based cloud management microcontroller, sensors, and an IP camera), and the real time lighting control function. | Solutions for the Smart Street Light Management System include employing adaptive controls to dynamically adjust lighting levels based on real-time conditions, optimizing energy consumption. Integrating motion sensors can further reduce energy waste by activating lights only when needed. |
| (Shaikh et al., 2018) | City pole controllability and enhance efficiency. | Smart Electric Pole System | LDR<br>Current<br>Accelerometer | It may face challenges in scaling up to cover an entire city or large urban. | Solutions for the Smart Electric Pole System could involve |
| | | | | High Initial Cost, Dependency on Connectivity. | implementing advanced sensor fusion algorithms to enhance accuracy in detecting issues like faults or tampering. |
| (Afifi et al., 2018) | Critical events messaging that occur in intermittent water distribution networks. | Monitoring and Burst Detection in Intermittent Water Distribution Networks | Point detection Sensor<br>Spot Type Sensor<br>Flow Meters | Built-in invasive Gauge pressure & External for flow or water quality Monitoring. | Some solutions which have to be implemented such as: Early detection of pipeline cracks, single leak detection, assumption of uniform pipeline material. |
| (Venkanna et al., 2018) | Low cost for implementation, Reduce the human fatigue, and time management. | Efficient Parking Slot Availability Detection System | Arduino (ATmega328)<br>(HC-SRO4)<br>Wi-Fi-ESP8266 | Limitations of the Efficient Parking Slot Availability Detection System include potential inaccuracies in detecting small vehicles or motorcycles, leading to misrepresented occupancy data. Adverse weather conditions may also impact sensor accuracy, affecting real-time information reliability. | Solutions for the Efficient Parking Slot Availability Detection System entail deploying ultrasonic or camera-based sensors to accurately monitor parking space occupancy in real-time. Utilizing data analytics, the system can predict peak parking times and guide drivers to available slots, reducing congestion. |
| (Pereira & Nagapriya, 2017) | Used by experts and doctors to provide an efficient solution. | Health Care Monitoring System | ECG Sensor<br>Wi-Fi-ESP8266 Module<br>AD8232 LM-35 | Limitations of the Health Care Monitoring System include potential challenges in ensuring consistent and reliable data transmission from wearable devices, affecting the accuracy of remote patient monitoring. The system's reliance on technology may lead to concerns over data privacy and security, especially if not properly safeguarded. | Solutions for the Health Care Monitoring System encompass implementing robust encryption and authentication protocols to safeguard patient data and ensure privacy. Utilizing data analytics and AI can enhance the accuracy of health insights derived from collected data, improving patient care. |
| (Sadhukhan, 2017) | Online booking, and checking vehicle’s from unfitting parking in the parking area. | E-Parking System | Arduino (MEGA 2560)<br>(HC-SRO4)<br>Wi-Fi Module IEEE 802.11 | Limitations of the E-Parking System include potential reliance on mobile app adoption, excluding individuals without smartphones or internet access. Accuracy in real-time parking space availability might be affected by delays in data updates or sensor malfunctions. | Solutions for the E-Parking System entail offering alternative methods of space availability notification, such as SMS alerts, for users without smartphones. Implementing redundant data communication channels can ensure more accurate and timely updates of parking availability information. |
| (Chaudhari & Bhole, 2018) | Locate empty bin in short path, Cost reduction, and Time management. | Solid Waste Collection System | Arduino (ATmega328P)<br>GPS (EM-506)<br>Wi-Fi-ESP8266 | Limitations of the Solid Waste Collection System include potential inaccuracies in fill-level sensors, leading to | Solutions for the Solid Waste Collection System involve optimizing collection routes through advanced algorithms, |
| | | | | inefficient collection routes and wasted resources. Network connectivity issues might hinder real-time data updates, affecting the system's responsiveness to changing waste levels. | reducing fuel consumption and emissions. Implementing smart bins with fill-level sensors enables efficient scheduling of collection services. |

In today's smart cities, cutting-edge technologies are revolutionizing urban services. The Smart Solid Waste Monitoring and Collection system uses IoT sensors to optimize waste collection, making it more efficient and environmentally friendly. With the Smart Street Light Management System, adaptive controls and IoT sensors ensure energy-efficient lighting and enhance safety. The Smart Electric Pole System provides a network of interconnected poles with advanced functionalities, contributing to improved urban services. Additionally, the Monitoring and Burst Detection in Intermittent Water Distribution Networks leverages technology to detect and address water leaks, ensuring a more reliable and sustainable water supply. These integrated smart solutions work together to create sustainable, efficient, and responsive urban environments.

The Efficient Parking Slot Availability Detection System optimizes urban mobility by providing real-time parking information, reducing congestion and improving user convenience. The Health Care Monitoring System leverages wearable devices and data analytics to enable remote patient monitoring, enhancing healthcare accessibility and individual well-being. The E-Parking System transforms parking management through digital solutions, yet it faces challenges related to equitable access and data accuracy. Meanwhile, the Solid Waste Collection System employs IoT sensors for efficient waste collection, although sensor reliability and public cooperation remain critical factors for its success. These systems collectively illustrate the potential of technology-driven solutions in addressing urban challenges, while also highlighting the need for careful implementation and user-centric design as shown in Table 8.

## RQ-06: What are the key principles and guiding practices for the successful implementation of IoT in smart cities?

**Collaboration and Stakeholder Engagement:** The successful implementation of IoT in smart cities necessitates collaboration among various stakeholders, including government agencies, private sector entities, technology providers, and citizens. Engaging all relevant parties in the planning, decision-making, and implementation process fosters inclusivity, transparency, and shared responsibility (Harmon et al., 2015).

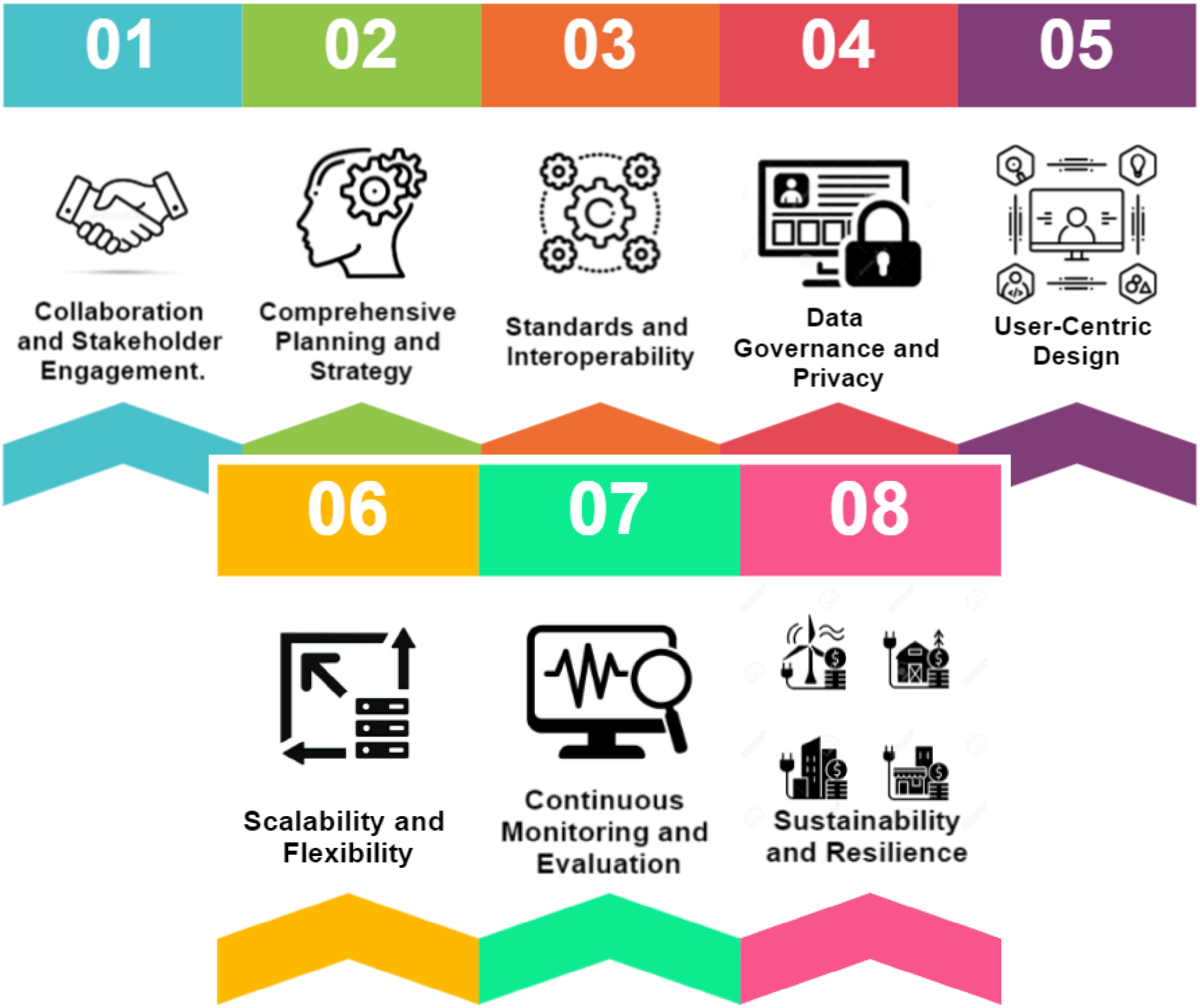

**Comprehensive Planning and Strategy:** Developing a comprehensive plan and strategy is crucial for successful implementation. This involves setting clear goals, defining the project's scope, and identifying desired outcomes. A well-defined roadmap helps prioritize initiatives, allocate resources effectively, and ensures alignment with the city's long-term vision (Ganchev et al., 2014).

**Standards and Interoperability:** Adhering to standards and promoting interoperability is essential. Utilizing open standards and protocols enables seamless integration and communication among different IoT devices, platforms, and systems. It ensures compatibility, scalability, and future-proofing of the smart city infrastructure (Ahlgren et al., 2016).

**Data Governance and Privacy:** Establishing robust data governance practices is critical for successful implementation. This includes defining data ownership, access controls, and data management protocols. Ensuring privacy protection, data security, and compliance with relevant regulations build trust among residents and stakeholders (Yang et al., 2019).

**User-Centric Design:** Placing residents and users at the center of the design process is crucial. Understanding their needs, preferences, and pain points helps develop user-friendly and intuitive IoT solutions. Incorporating user feedback and iterative design processes enhance user adoption, satisfaction, and the overall success of the implementation (Guo et al., 2018).

**Scalability and Flexibility:** Planning for scalability from the outset is important. Anticipating future growth and considering the ability to expand the infrastructure, networks, and services ensures the long-term viability of the IoT implementation. Embracing flexible architectures and modular approaches allow for easy integration of new technologies and adaptation to evolving needs (Zedadra et al., 2019).

**Continuous Monitoring and Evaluation:** Implementing mechanisms for continuous monitoring and evaluation is essential. Regular assessment of the performance, impact, and effectiveness of the IoT solutions helps identify areas for improvement, measure progress towards goals, and make informed decisions for optimization and refinement (Andrade et al., 2020).

**Sustainability and Resilience:** Considerations for sustainability and resilience are vital. Implementing energy-efficient solutions, utilizing renewable energy sources, and optimizing resource utilization contribute to long-term sustainability. Additionally, integrating resilience features such as backup systems, disaster management capabilities, and redundancy ensures the robustness of the smart city infrastructure (Ramirez Lopez & Grijalba Castro, 2020).

# Conclusion:

The selection process began with an extensive search using various relevant terminologies associated with IoT in Smart Cities. The results were then carefully analyzed, and the search was concluded in July 2023, ensuring that studies carried out after this date were not included. The Web of Science core collection was utilized for the analysis, and out of 20,653 publications, 55 articles were selected for further examination.

The findings revealed that almost all of the selected articles were published in reputable journals, with some research paper presented at a conference. The two primary types of studies adopted in these articles were "Solution Proposal" and "Evaluation Research." Furthermore, the majority of the chosen research was evidence-based, indicating that they could potentially yield significant advantages in the context of IoT in smart cities. Through accurately following this selection process, the review ensures the inclusion of high-quality and relevant studies that contribute valuable insights into the applications and impact of IoT in Smart Cities.

However, in this research we focused on the pivotal role of IoT in smart cities, with a specific emphasis on smart home, smart infrastructure, and smart industry. Our objective was to explore how these components, when, can enhance urban efficiency and sustainability while also addressing the challenges associated with their individual implementations. Through a comprehensive analysis, we identified the security concerns in each domain, highlighting the need for robust measures to protect data and infrastructure. As a result, we successfully integrated these IoT components into a unified framework, presenting a holistic approach to building smart cities of the future. This research emphasizes the transformative potential of IoT technology and offers valuable insights for creating interconnected, efficient, and resilient urban ecosystems.

## Limitations of above studies:

"**Internet of things: A survey on enabling technologies, protocols, and applications**" survey provides a comprehensive overview, but it might lack in-depth analysis of specific enabling technologies and protocols, leaving out some detailed insights. "**Security analysis of network anomalies mitigation schemes in IoT networks**" study focuses on network security, but it may not address all possible mitigation schemes and might not consider emerging threats or recent advances in the field. "**IoT in smart cities: a survey of technologies, practices, and challenges",** while offering valuable insights, the survey may not cover every aspect of IoT implementation in smart cities, potentially overlooking specific challenges faced by different urban environments.

"**Networking architectures and protocols for smart city systems**" study explores networking solutions, but it may not delve deep into the practical implementation challenges faced by smart cities and the scalability of proposed architectures. "**Investigating emerging technologies role in smart cities solutions**", although highlighting emerging technologies, the investigation might not extensively evaluate their real-world applications and may not address their limitations in specific use cases. "**Security and privacy for green IoT-based agriculture: Review, blockchain solutions, and challenges**", while providing valuable insights into agriculture IoT security, the study may not comprehensively cover all potential blockchain solutions and may not fully address privacy concerns in the context of green agriculture IoT.